\documentclass{article}
\usepackage{authblk}
\usepackage{amssymb}
\usepackage{amsmath}
\usepackage{amsthm}
\usepackage{booktabs}
\usepackage{multirow}
\usepackage{longtable}
\usepackage{calc}
\usepackage{array}
\usepackage{comment}
\usepackage{authblk}
\usepackage[a4paper]{geometry}
\usepackage[round]{natbib}
\usepackage[dvipsnames]{xcolor}
\usepackage[colorlinks,linktoc=all,
             citecolor= Cerulean!85!green!90!black,
             linkcolor=YellowOrange,
             urlcolor=RubineRed!85!black,
             ]{hyperref}
\usepackage{graphicx}

\title{Toward a Team of AI-made Scientists for Scientific Discovery from Gene Expression Data}
\author{
    Haoyang Liu\textsuperscript{1}, 
    Yijiang Li\textsuperscript{2}, 
    Jinglin Jian\textsuperscript{1}, 
    Yuxuan Cheng\textsuperscript{3}, 
    Jianrong Lu\textsuperscript{4}, 
    Shuyi Guo\textsuperscript{1}, 
    Jinglei Zhu\textsuperscript{1}, 
    Mianchen Zhang\textsuperscript{1}, 
    Miantong Zhang\textsuperscript{1}, 
    Haohan Wang\thanks{Corresponding author. Email: haohanw@illinois.edu} \textsuperscript{1}\\
    \small \textsuperscript{1}University of Illinois at Urbana-Champaign\\
    \small \textsuperscript{2}University of California, San Diego\\
    \small \textsuperscript{3}Huazhong Agricultural University\\
    \small \textsuperscript{4}Huazhong University of Science and Technology
}
\date{}

\begin{document}

\maketitle

\begin{abstract}
Machine learning has emerged as a powerful tool for scientific discovery, enabling researchers to extract meaningful insights from complex datasets. For instance, it has facilitated the identification of disease-predictive genes from gene expression data, thereby improving risk stratification, early diagnosis, and treatment selection. However, the traditional process for analyzing such datasets demands substantial human effort and expertise for the data selection, processing, and analysis. To address this challenge, we introduce a novel framework, a Team of AI-made Scientists (TAIS), designed to streamline the scientific discovery pipeline. TAIS comprises simulated roles, including a project manager, data engineer, and domain expert, each represented by a Large Language Model (LLM). These roles collaborate to replicate the tasks typically performed by data scientists, with a specific focus on identifying disease-predictive genes. Furthermore, we have curated a benchmark dataset to assess TAIS's effectiveness in gene identification, demonstrating our system's potential to significantly enhance the efficiency and scope of scientific exploration. Our findings represent a solid step towards automating scientific discovery through large language models.
\footnote{Code for a more recent version of our system is available at \url{https://github.com/Liu-Hy/GenoMAS}.}
\end{abstract}

\section{Introduction}
\label{intro}
In the late 1990s, Netherlands Cancer Institute scientists applied machine learning and discovered 70 predictive genes for cancer spreading \citep{van2002gene}, leading to the creation of MammaPrint, a diagnostic tool for assessing cancer risk and guiding early-stage treatment \citep{mook2007individualization,brandao2019mammaprint}. MammaPrint sparked a billion-dollar industry and aided numerous women in cancer diagnosis and treatment. This remarkable success highlights the immense potential of machine learning in analyzing gene expression data. The availability of gene expression databases, such as the Cancer Genome Atlas (TCGA) \citep{tomczak2015cancer} and the Gene Expression Omnibus (GEO) \citep{clough2016gene}, 
opens up vast opportunities for scientists to explore disease-related genes. With these resources, 
researchers can potentially uncover new genes important in disease development, potentially helping a wider spectrum of people suffering from various health conditions.

Furthermore, the emerging field of personalized medicine
\citep{hamburg2010path,chan2011personalized}
highlights the need for a more careful analysis. 
It is important 
to recognize that key genes linked to diseases may vary under different physical conditions. 
Therefore, studies should consider a diverse set of factors like age, gender, and co-occurrences of other diseases. 
Incorporating these conditions into research designs 
can help us gain a more comprehensive understanding of the underpinnings of these diseases. 

This approach holds the promise of helping a broad range of patients by understanding diseases and tailoring treatments to individual needs. However, it also comes with significant challenges, such as navigating vast gene expression datasets \citep{hulsen2023bigdata} and addressing potential confounding factors. Additionally, researchers need to possess technical proficiency in coding, data processing, and analysis, requiring a blend of scientific knowledge and advanced analytical skills. These complexities highlight the difficulties in leveraging data analysis to benefit patients.


Leveraging Large Language Models (LLMs) \citep{patil2024review} as agents \citep{guo2024largelanguagemodelbased}, this paper proposes a Team of AI-made Scientists (TAIS) to automatically simulate researchers' work. TAIS agents will execute tasks like dataset selection, preprocessing, confounder factor correction, condition prediction, and analysis to identify disease-predictive genes under various conditions.


Given the pioneering nature and complexity of our aim, We then establish a gold standard benchmark with datasets comprising 457 disease-condition pairs to assess our TAIS method's performance. This benchmark involves manual selection, analysis, and processing of datasets, as well as writing and executing code to identify predictive genes for disease status under various conditions.


Our evaluation demonstrates that TAIS can effectively perform intricate data analysis on genetic datasets, and its performance can be further improved through the iterative process of collaboration among agents. Our case study reveals that the genes identified by our agents are corroborated by biomedical research.

In summary, the contribution of our paper is as follows. 
\begin{itemize}
    \item We introduced the Team of AI-made Scientists (TAIS) system, an agent system simulating scientific research activities for analyzing genes predictive of disease under various conditions. 
    \item Beyond standard data analysis tasks (i.e. data processing, analysis), we introduced crucial steps like confounding factor correction to minimize false discoveries and two-step regression to account for missing conditions.
    \item We developed a benchmark to evaluate our TAIS method, simulating human scientists' data analysis process and documenting errors for future reference.
\end{itemize}

\section{Related Work}
\label{related}
\begin{figure*}
    \centering
    \includegraphics[width=0.8\textwidth]{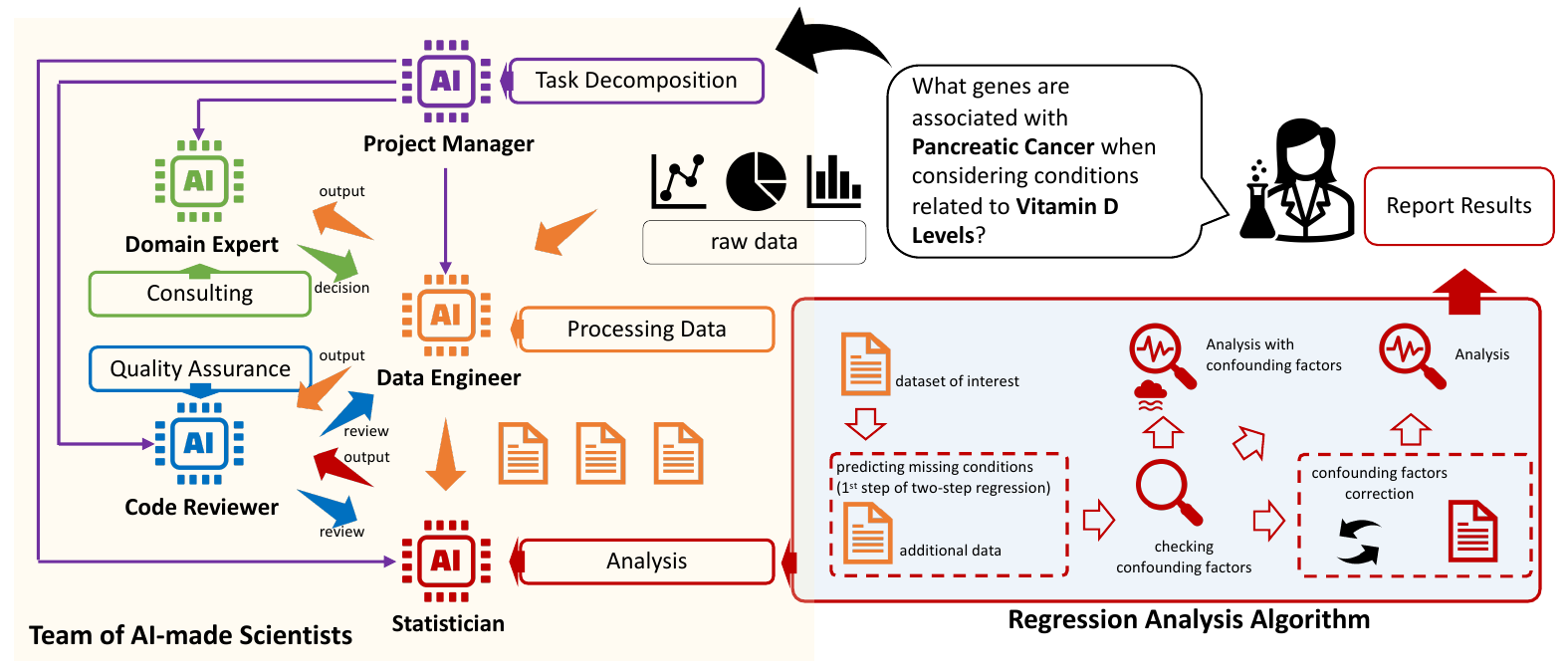}
    \caption{The overview of the Team of the AI-made Scientists (TASI). The illustration starts from the top right corner where the user uses the system. The question goes to the project manager. The project manager further decomposes the tasks and assigns tasks to different AI-made scientists, illustrated in the yellow area. The blue area shows the details of how the statistician analyzes the data. }
    \label{fig:main}
\end{figure*}
\subsection{Pipelines to Identify Genes Predictive of Disease Status}

The pipeline to analyze gene expression data with ML begins with dataset selection (e.g., GEO \citep{clough2016gene} or TCGA \citep{tomczak2015cancer}) and preprocessing: cleaning, handling missing values \citep{abusamra2013comparative}, removing empty records, and excluding unrecorded genes \citep{khondoker2006statistical}. One then fits regression models to identify genes predictive of disease status \citep{ghosh2005classification,wu2009genome}, often using Lasso for sparsity \citep{tibshirani1996regression}. To reduce bias, pipelines correct confounding/batch effects \citep{leek2010tackling,bruning2016confounding, yu2006unified}. Later work further integrates covariates (demographics, comorbidities) to enable precision analyses \citep{yang2023overview, kyalwazi2023race, rosenquist2023novel}.
More recently, several studies refined these pipelines along complementary axes. Comparative work assessed how preprocessing choices impact cross-study generalization for transcriptomic prediction, underscoring the importance of harmonized workflows \citep{mize2024rnaseq}. Methodologically, deep learning and graph-based models improved disease–gene association discovery from omics graphs and sequence-derived representations \citep{saadat2024dnalang}. At the interface of genomics and pathology, recent reviews synthesize advances in deep learning that connect molecular profiles with histopathology, informing pipeline design for robust biomarker discovery \citep{unger2024deeplearning}.

\subsection{Task Solving via LLMs as Agents}
LLMs show strong general capabilities \citep{wang2023survey, openai2023gpt4, touvron2023llama, touvron2023llama2}. Work on reasoning and acting \citep{wang2022towards, wang2022self, hao2023reasoning, yao2022react} popularized CoT \citep{wei2022chain} with goal decomposition \citep{zheng2023synapse, feng2023towards, wang2022towards, ning2023skeleton}. Multi-agent systems further amplify problem solving \citep{wang2023unleashing, talebirad2023multiagent, du2023improving, wang2023adapting, yang2023autogpt, dong2023selfcollaboration}, and role-based frameworks like MetaGPT operationalize collaboration \citep{hong2023metagpt, qian2023communicative}. As evaluation has matured in 2024, new benchmarks probe complementary dimensions: dynamic multi-agent competence (LLMArena) \citep{chen2024llmarena}, cooperation/competition (BattleAgentBench) \citep{wang2024battleagentbench}, safety risks in interactive settings (Agent-SafetyBench) \citep{zhang2024agentsafetybench}, and progress tracking via modular tasks and metrics (AgentQuest) \citep{gioacchini2024agentquest}. Beyond enabling software tasks, recent efforts target end-to-end science: the AI Scientist proposes automated idea generation, experimentation, and paper writing \citep{lu2024aicscientist}. Complementary, forecasting pipelines use evolving knowledge graphs to predict emergent impactful directions \citep{gu2024forecasting}. In contrast to domain-specific finetuning in chemistry/biotech/medicine \citep{bran2023chemcrow, guo2023can, richard2024chatnt, tang2024mimir}, we leverage off-the-shelf LLMs and coordinate a team for genomics.

\section{Method}
\label{method}

\subsection{System overview: a lightweight role-driven team}
\label{sec:tais}

TAIS organizes a small set of specialized agents into a two-stage pipeline for gene expression analysis: data preparation followed by 
regression-based association testing. The team comprises five roles with minimal but complementary responsibilities: 
\textbf{Project Manager} (coordinator) parses the user query (trait and optional condition), scopes required datasets (e.g., TCGA, GEO), 
and schedules two sequential stages with checkpoints. \textbf{Data Engineer} implements dataset-specific preprocessing code. 
\textbf{Statistician} runs regression to identify trait-associated genes while accounting for confounding. \textbf{Domain Expert} acts 
as a biomedical consultant for decisions that hinge on domain knowledge (e.g., cohort inclusion, clinical variable parsing, gene symbol 
normalization). \textbf{Code Reviewer} audits generated code for executability and instruction conformance.

Execution is intentionally simple. The Project Manager issues stage descriptions and acceptance criteria; the Data Engineer and 
Statistician write short code segments, execute them, and submit outputs and logs. The Code Reviewer provides bounded feedback (a small, 
fixed number of review rounds) when code fails or drifts from instructions. The Domain Expert is queried only at decision points 
requiring biomedical judgment (e.g., mapping histology strings to labels, platform-specific gene identifier handling). This lightweight 
design avoids heavy orchestration machinery while still enforcing basic quality control.

\begin{figure}[htbp]
\centering
\begin{minipage}[t]{0.48\textwidth}
\centering
\includegraphics[width=1.0\textwidth]{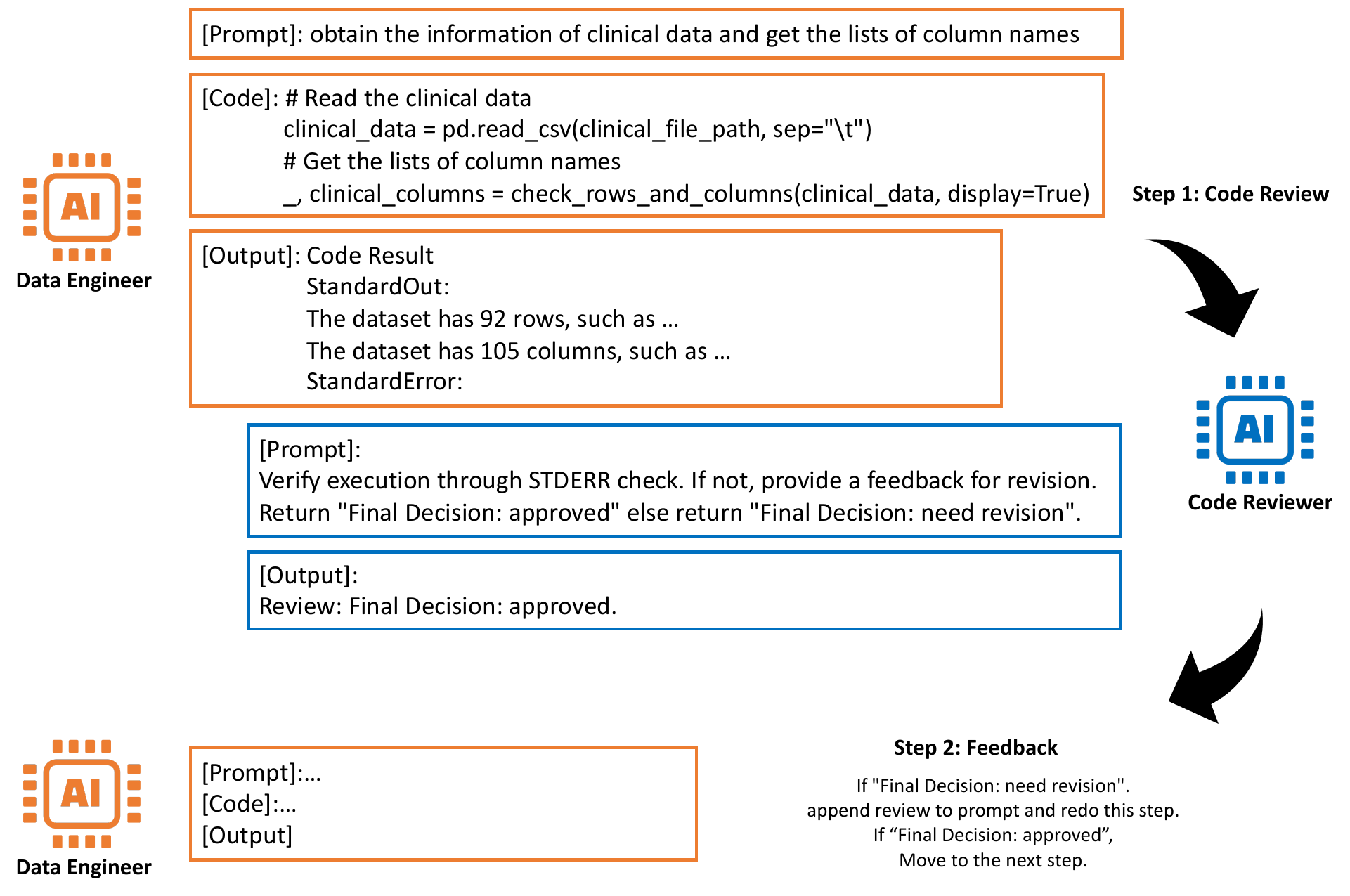}
    \caption{Write--run--audit loop between the Data Engineer and Code Reviewer.}
    \label{fig:reviewer}
\end{minipage}
\begin{minipage}[t]{0.48\textwidth}
\centering
\includegraphics[width=1.0\textwidth]{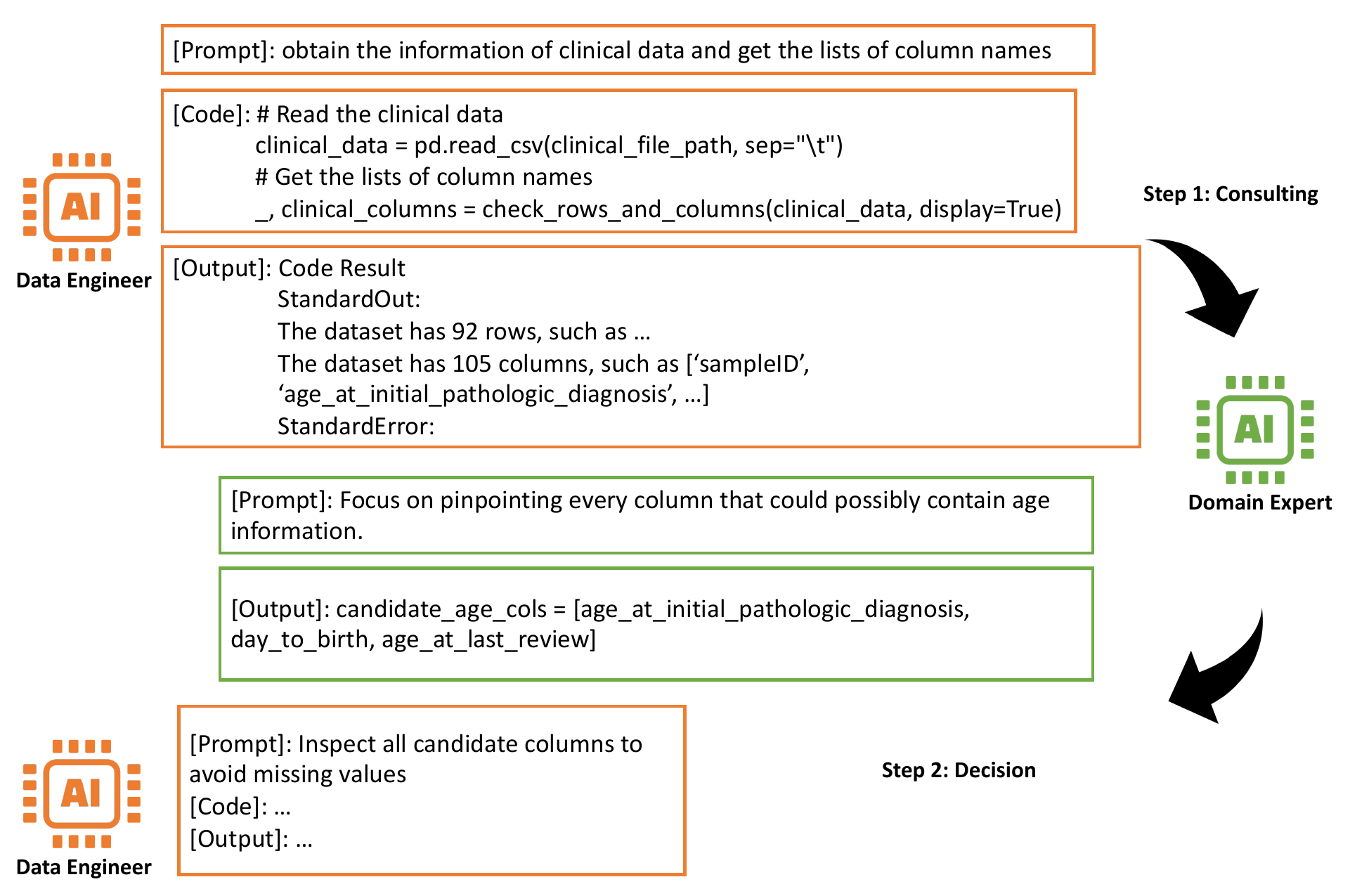}
    \caption{Consultation between the Data Engineer and Domain Expert on biomedical decisions.}
    \label{fig:domain}
\end{minipage}
\end{figure}

For brief role descriptions, see Appendix~\ref{sec: role_details}. An overview schematic is shown in Figure~\ref{fig:main}. In TAIS, 
three agents (Data Engineer, Domain Expert, Code Reviewer) collaborate on preprocessing, while the Statistician and Code Reviewer handle 
regression.

\subsection{Collaboration among AI Scientists}
\label{sec:collaboration}

We employ two interaction patterns:

\textbf{Write--Run--Audit (program-and-review).} For any step that generates code, the authoring agent (Data Engineer or Statistician) 
executes the code and submits the snippet, stdout/stderr, and step instruction to the Code Reviewer (Figure~\ref{fig:reviewer}). The 
reviewer checks two things: (i) whether the code runs without errors and (ii) whether it follows the instruction and acceptance criteria. If 
rejected, the author revises and re-runs. The loop is bounded by a small review budget to limit latency and overfitting to feedback.

\textbf{Consultative Coding.} When a step depends on biomedical knowledge (e.g., extracting clinical labels from free-text metadata, 
merging gene identifiers across platforms), the Data Engineer requests guidance from the Domain Expert (Figure~\ref{fig:domain}). The 
expert returns concise, actionable advice or pseudo-code that the engineer turns into executable code. If execution fails, the 
consultation can repeat within the same step until the review budget is exhausted.

\paragraph{End-to-end flow.} In the end-to-end analysis flow, given a query (trait and optional condition), the Project Manager (i) locates candidate cohorts, (ii) 
chooses stage order and checkpoints, and (iii) starts preprocessing. The Data Engineer performs file parsing, sample filtering, clinical 
feature extraction, gene symbol normalization, and normalization of expression values. The outputs are matrix-shaped tables with aligned 
sample identifiers and optional condition columns. After the audit passes, the Statistician selects a regression recipe 
(Section~\ref{sec:regression_summary}) based on basic diagnostics and produces a ranked list of genes with effect estimates. All steps 
are auditable and rerunnable with the same inputs and random seeds.

\subsection{Regression on gene expression with basic confounding control}
\label{sec:regression_summary}

Analyzing gene expression data to identify significant gene factors, considering confounding variables, involves a comprehensive statistical approach due to the data's high-dimensionality and heterogeneity. 
We summarize the statistical backbone used by the Statistician.

To address the challenges of variable selection in high-dimensional data, Lasso regression is employed to isolate influential genes. The detection of confounding factors is informed by analyzing the eigenvalue gaps of the covariance matrix of input features. 
Confounding factors, if present, necessitate either regression with confounding factor correction or regression after confounding factor adjustment, employing a linear mixed model (LMM)~\citep{yu2006unified,lippert2011fast,wang2022trade} or a data transformation approach respectively.

Further, to incorporate additional conditions such as age and gender into the analysis, 
the residualization approach is employed to account for the effect of the condition. 
For cases lacking direct condition data, a two-step regression strategy is adopted. This involves using common genes between datasets to estimate missing conditions, thus enabling the integration of trait and condition data for comprehensive analysis.


\section{Benchmark Creation}
\label{Benchmark_generation}
To streamline the evaluation of our TAIS approach and promote future research with large language models in genetic data analysis, we developed the Genetic Question Exploration (GenQEX) dataset. This benchmark dataset consists of 457 carefully selected questions, complete with a comprehensive gold standard that includes genetic datasets from public sources, preprocessing and regression analysis code, and the corresponding results. Here, we outline the process of creating this benchmark.

\paragraph{Question Generation}
A computational biology researcher identified a list of key biomedical entities related to genetics research or public health, resulting in 65 traits classified into 9 categories. These traits were paired with either another trait or demographic attributes like ``age'' or ``gender'', generating 4556 possible pairs. These pairs were designed to pose questions in the format: \emph{``What are the significant genes related to the trait when considering the influence of the condition?''} Then, we used a set of inclusion and exclusion criteria (Appendix~\ref{sec:criteria}) and ranked the trait-condition association of pairs based on the Jaccard similarity of related genes from the NCBI Gene database, to identify 457 pairs that are of most scientific interest, which form our benchmark's question set. Details on these pairs are available in Appendix~\ref{sec:question_pairs}.

\paragraph{Input Dataset}
Gene expression and clinical data were obtained from the Gene Expression Omnibus (GEO) \citep{clough2016gene} and the Cancer Genome Atlas (TCGA) \citep{tomczak2015cancer} through the Xena platform \citep{goldman2020visualizing}. Gene symbols related to traits were sourced from the NCBI Gene database \citep{brown2015gene}. For more information on these data sources, see Appendix~\ref{sec:datasource}.

To enhance the evaluation of our TAIS method and facilitate future research utilizing large language models in genetic data exploration, we introduce a benchmark dataset comprised of 457 meticulously formulated questions, alongside a gold standard for resolving each, which encompasses genetic datasets sourced from public repositories, preprocessing and regression analysis code, and corresponding results. Our benchmark is named as the Genetic Question Exploration (GenQEX) dataset. This section delineates the benchmark's development process. 
\paragraph{Question Generation}
A computational biology researcher curated a list of significant biomedical entities, encompassing human traits or diseases pivotal to genetics research or public health. After a thorough manual curation, a diverse list of 67 traits across 9 primary categories was compiled. Subsequently, each trait was paired with a condition---either another trait from the list or one of the demographic attributes ``age'' or ``gender''---resulting in 4556 potential pairs. These pairs were designed to pose questions in the format: \emph{``What are the significant genes related to the trait when considering the influence of the condition?''} To sift through these pairs for scientifically relevant queries, we employed specific inclusion and exclusion criteria.
The pairs were then ranked based on the Jaccard similarity between the related genes of the trait and the condition, derived from the NCBI Gene database, to identify pairs where the trait is implicated with the condition. This process yielded 457 pairs, which collectively form our benchmark's question set. For a comprehensive list of these pairs please see Appendix~\ref{sec:question_pairs}.
\paragraph{Input Dataset}
To address the formulated research questions, gene expression and corresponding clinical data were procured from renowned public databases: The Cancer Genome Atlas (TCGA) via the Xena platform, and the Gene Expression Omnibus (GEO). Additionally, domain knowledge regarding gene symbols associated with traits was sourced from the NCBI Gene database. Please refer to Appendix~\ref{sec:datasource} for an introduction about these data sources.

\paragraph{Manual Curation}
A dedicated team of four researchers within our group undertook the task of curating the question list and extracting relevant input data from public sources. 
A subsequent phase of manual curation involved nine computer science researchers who meticulously developed the gold standard, comprising preprocessing and regression analysis code, and the outcomes. Equipped with detailed instructions and the solutions to example questions, these researchers crafted the gold standard for all listed traits over three weeks. A computational biologist provided ongoing review to ensure the rigor of the example code and instructions.



\section{Experiment}
\label{Experiment}
\subsection{Experiment Setting}

We evaluate TAIS on our benchmark of gene-trait association tasks using five metrics: Success Rate (SR), Precision, Recall, F$_1$, and Jaccard. We consider three complementary protocols to isolate contributions of each stage:
(1) end-to-end: TAIS performs both preprocessing and regression;
(2) regression-only: we replace the inputs with gold-standard preprocessed data to assess the Statistician in isolation;
(3) preprocessing-only: we feed data preprocessed by TAIS into a gold-standard regression script to assess the Data Engineer in isolation.
For stages involving code generation, we vary the review budget (maximum number of write-run-audit rounds) as defined in Section~\ref{sec:collaboration}.

\subsection{Main Results}
We first present the end-to-end results in Section~\ref{sec:main_results}. We then present evaluation of the regression-only setting in Section~\ref{sec:regression} and the preprocessing-only setting in Section~\ref{sec:data_preprocessing}.

\subsubsection{Performance of TAIS System}
\label{sec:main_results}
Table~\ref{tab1} summarizes end-to-end performance. Overall, TAIS attains SR 69.08\%, Precision 33.91\%, Recall 31.70\%, F$_1$ 30.27\%, and Jaccard 21.13\%. The single-step setting is substantially easier than the two-step setting (F$_1$ 45.05\% vs. 19.21\%), reflecting the difficulty of estimating or integrating missing conditions across cohorts. These results are consistent with our design choices: a lightweight team with bounded review budgets yields reasonable performance, but struggles most when preprocessing must infer conditions.

\begin{table}[ht]
    \centering
    \footnotesize
    \caption{Performance of TAIS System on our benchmark. We provide performance on single-step and two-step regression tasks respectively. }
    \label{tab1}
    \setlength{\tabcolsep}{4mm}{
    \begin{tabular}{@{}lccccc@{}}
        \toprule
         & Success Rate (\%) & Precision (\%) & Recall (\%) & F$_1$ (\%) & Jaccard (\%) \\
        \midrule
        Single-step  &  71.27 & 48.35 & 43.84 & 45.05 & 30.15 \\
        Two-step  & 67.43 & 23.01 &  22.55 & 19.21 & 14.33 \\
        Overall   & 69.08 & 33.91 & 31.70 & 30.27 & 21.13 \\
        \bottomrule
    \end{tabular}}
\end{table}

\subsubsection{Performance of Regression}
\label{sec:regression}
To evaluate the Statistician, we use gold-standard preprocessed datasets as inputs and vary the review budget for program-and-review. Results in Table~\ref{tab2} show that code review markedly improves performance: overall F$_1$ increases from 55.95\% (0 reviews) to 80.12\% (1 review) and 89.30\% (2 reviews). Gains are consistent across single-step (F$_1$ 58.61\% to 94.74\%) and two-step (53.77\% to 85.20\%), indicating that most of the statistical errors are correctable through bounded iteration. The Statistician's logic is thus strong when provided with clean inputs.

\begin{table}[ht]
    \centering
    \caption{Performance of TAIS on regression analysis of our benchmark}
    \label{tab2}
    \resizebox{0.9\textwidth}{!}{
    \setlength{\tabcolsep}{4mm}{
    \begin{tabular}{@{}lcccccc@{}}
        \toprule
         & Budget (\# reviews) & Success Rate (\%) & Precision (\%) & Recall (\%) & F$_1$ (\%) & Jaccard (\%) \\
        \midrule
        \multirow{3}{*}{Single-step}  & 0 & 62.56 & 58.85 & 58.45 & 58.61 & 43.13 \\
        &  1 & 87.97 & 87.02 & 84.84 & 85.62 & 77.54 \\
        &  2 & 97.45 & 95.88 & 93.71 & 94.74 & 89.38 \\
        \midrule
        \multirow{3}{*}{Two-step} & 0 & 58.80 & 56.22 & 51.88 & 53.77 & 38.59 \\
          &  1 & 83.87 & 78.31 & 75.19 & 74.92 & 63.24 \\
            & 2 & 91.33 & 85.88 & 84.44 & 85.20 & 76.11 \\
        \midrule
        \multirow{3}{*}{Overall} & 0 & 60.42 & 57.25 & 54.71 & 55.95 & 41.68 \\
           &  1 & 85.63 & 81.05 & 79.34 & 80.12 & 69.39 \\
           &  2 & 93.96 & 90.18 & 88.43 & 89.30 & 81.83 \\
        \bottomrule
    \end{tabular}}}
\end{table}

The sharp jump from 0 to 1 review highlights the value of the audit loop for catching implementation drift and minor numerical issues; the second review yields diminishing but still notable returns. Since inputs are fixed and clean, the remaining gap to end-to-end performance mainly stems from preprocessing quality.

\begin{table}
    \centering
    \caption{Performance of TAIS on data preprocessing}
    \label{tab3}
    \resizebox{0.9\textwidth}{!}{
    \setlength{\tabcolsep}{4mm}{
    \begin{tabular}{@{}lcccccc@{}}
        \toprule
         & Budget (\# reviews) & Success Rate (\%) & Precision (\%) & Recall (\%) & F$_1$ (\%) & Jaccard (\%) \\
        \midrule
        \multirow{3}{*}{Single-step} & 0 & 36.62 & 23.72 & 22.81 & 21.93 & 16.97 \\
        & 1 & 68.98 & 40.95 & 39.17 & 40.03 & 25.63 \\
        & 2 & 78.95 & 50.36 & 46.89 & 47.17 & 32.07 \\
        \midrule
        \multirow{3}{*}{Two-step} & 0 & 33.48 & 10.45 & 9.62 & 8.07 & 6.19 \\
          & 1 & 59.54 & 22.09 & 21.95 & 18.04 & 14.28 \\
          & 2 & 70.30 & 26.24 &  26.76 & 22.33 & 16.66 \\
        \midrule
        \multirow{3}{*}{Overall} & 0 & 34.83 & 16.16 & 15.29 & 14.41 & 10.83 \\
            & 1 & 63.59 & 30.20 & 29.35 & 27.50 & 19.16 \\
            & 2 & 74.02 & 36.61 & 35.42 & 33.01 & 23.29 \\
        \bottomrule
    \end{tabular}}}
\end{table}

\subsubsection{Performance of Data Preprocessing}
\label{sec:data_preprocessing}
To assess the Data Engineer, we execute gold-standard regression code on TAIS-preprocessed outputs. Table~\ref{tab3} shows that review budget strongly impacts quality: overall F$_1$ improves from 14.41\% (0 reviews) to 33.01\% (2 reviews), with corresponding SR increasing from 34.83\% to 74.02\%. Single-step tasks benefit the most (F$_1$ from 21.93\% to 47.17\%), whereas two-step tasks remain challenging (8.07\% to 22.33\%), underscoring the difficulty of extracting and harmonizing condition signals from heterogeneous metadata.

Comparing Tables~\ref{tab2} and \ref{tab3}, the dominant bottleneck is preprocessing: even with gold-standard regression, performance remains far below the regression-only setting. Nevertheless, program-and-review substantially reduces errors and almost doubles F$_1$ on two-step tasks, indicating that modest iteration and targeted feedback are highly valuable for data preparation in this domain.

\section{Case Study}
\label{Case}
To offer a more direct understanding of 
the performances of our TAIS system, 
we detail a case study of one particular research question here. 
When our system is asked with 
``What genes are associated with Pancreatic Cancer when considering conditions related to Vitamin D Levels?''
Our system identified 20+ genes
with a disease (Pancreatic Cancer) prediction cross-validation accuracy 
of $80\%$. 

The top five genes identified are 
\textit{SLC11A1}, \textit{SOCS1}, \textit{CD207}, 
\textit{LILRB3}, and \textit{SPA17}.
Out of these five genes, four are implicated with 
Pancreatic Cancer when considering the interaction 
with Vitamin D Levels. 

\textit{SLC11A1} has been implicated in the host's response to pathogens and may also play a role in inflammatory diseases \cite{awomoyi2007human}. Given that inflammation is a known risk factor for pancreatic cancer, and vitamin D is involved in modulating inflammatory responses \cite{colotta2017modulation}, \textit{SLC11A1} could be a link between vitamin D levels and inflammation-related pancreatic cancer risk.

\textit{SOCS1} is a critical regulator of cytokine signaling pathways, including those involved in immune responses and inflammation \cite{ying2019socs1}. Vitamin D is known to modulate immune function \cite{baeke2010vitamin} and inflammation \cite{colotta2017modulation}, suggesting that SOCS1 could be part of the pathway through which vitamin D influences pancreatic cancer risk or progression.

\textit{CD207} is a C-type lectin receptor expressed on Langerhans cells, which are involved in immune responses in the skin but might also play roles in other types of immune responses. While the direct link between CD207, vitamin D, and pancreatic cancer is less clear, the potential connection might relate to the broader immune modulation by vitamin D and how it could affect cancer immunosurveillance.

\textit{LILRB3} is involved in the regulation of immune responses, including the inhibition of various cell signaling pathways. Vitamin D has been shown to influence the immune system \cite{baeke2010vitamin}, and alterations in LILRB3 function could potentially affect how the immune system responds to cancer cells in the context of varying vitamin D levels.

\textit{SPA17} is known for its expression in reproductive tissues and certain cancers. It may play a role in cancer cell mobility and immune evasion. Given vitamin D's effects on immune function \cite{baeke2010vitamin}, there could be a link between vitamin D levels and the immune response to pancreatic cancer cells expressing SPA17, impacting the disease's progression or response to therapy.

\section{Conclusion}
\label{con}
We present a transformative approach to streamline the scientific discovery process through the development of a Team of AI-made Scientists (TAIS). TAIS comprises various roles, such as Project Manager and Domain expert, Each simulated by a Large Language Model (LLM). This team collaborates to execute tasks traditionally performed by data scientists, such as data preprocessing and analysis, with a focus on identifying genes predictive of disease status. To assess the efficacy of TAIS, we curated a benchmark dataset specifically for the evaluation of its performance in this domain. Our findings demonstrate a promising direction in automating the scientific discovery process, highlighting the potential of TAIS to reduce the human effort and technical expertise required in the analysis of scientific data.

\newpage 
\section*{Impact Statement}
This paper presents work whose goal is to advance the field of scientific discover in genomics. We hope our work can provide insights for domain experts such as geneticists and help them diagnose multiple diseases in a more personalized manner. We suggest that our model should be used under human supervision to ensure a perfect result. There are many other potential societal consequences of our work, none which we feel must be specifically highlighted here.

\bibliography{ref}
\bibliographystyle{abbrvnat}

\newpage
\appendix
\onecolumn
\label{supplementary_materials}
\section{Details of the roles}
\label{sec: role_details}
Below we introduce the roles of different agents in our Team of AI Scientists (TAIS).

\textbf{Project Manager} First of all, a Project Manager is the initiator of the TAIS with an overall view of the scientific problem (e.g. specifications and objectives) and the full knowledge of all available AI-made scientists (e.g. their capabilities). 
The project manager will first decompose the problem into several sub-problems based on the capabilities of available agents. For instance, the Project Manager is aware that the two datasets, i.e. TCGA and GEO, are present in the problem. Also, in the TAIS team, the Data Engineer is capable of data preprocessing and the Statistician is capable of regression analysis. Thus the Project Manager decomposes the the task of identifying genes predictive of the disease into three sub-problems, i.e. preprocessing of TCGA, preprocessing of GEO and regression analysis,  as illustrated in Figure~\ref{fig:main}. 

After problem decomposition, the Project Manager will then recruit AI-made scientists and assign them to each sub-problem. As illustrated in Figure~\ref{fig:main}, Data Engineer, Domain Expert and Code Reviewer are assigned to perform the two preprocessing tasks and the Statistician along with the Code Reviewer are assigned to the regression analysis task. We detail the collaboration among agents in Section \ref{sec:collaboration}. There are two types of Statisticians, capable of performing single-step and two-step regression respectively, as detailed later. The Project Manager will decide which Statistician agent is recruited for the regression analysis based on the condition variable, i.e. if the condition is Age / Gender, then recruit a single-step Statistician, otherwise a two-step Statistician.

\textbf{Data Engineer}
A Data Engineer is designed with skills for data engineering, i.e. data analysis, code writing, and execution. In TAIS, the Data Engineer is assigned the task of preprocessing datasets, i.e. TCGA and GEO datasets. Specific context regarding the datasets is first given to the Data Engineer including the path to the raw dataset directory, the overall research question, the trait of interest, and related function tools. Then the Data Engineer will follow the instructions to perform the preprocessing process.

Moreover, the Data Engineer is capable of coding in an interactive environment. 
This is particularly important because each 
 processing step is conditioned not only the previous step 
but also the specific data structures and the genomics information. 
Thus, the Data Engineer will have to execute the code at each step, check the data, and consult the Domain Expert before entering the next step. 
Thus, we enable the Data Engineer the ability to execute code at any step which in turn provides feedback to the coding. 

To facilitate the Data Engineer in process the data in an interactive environment, we empower the agent with code execution ability with multi-step instructions and assistance from Domain Expert. As illustrated in Figure~\ref{fig:domain}, at each step, Data Engineer will program following the corresponding instruction, execute the code, gather the output and asks the Domain Expert for domain information which is used for prompting the code for the next step. 


\textbf{Domain Expert}
A Domain Expert is a professional who conducts scientific investigations and experiments to understand and improve human health. 
Gene expression datasets are filled with biomedical terminologies, customary abbreviations, and technical descriptions of the data collection process, which are often only understandable by experts in related fields. These experts possess the knowledge and experimental techniques required to understand the samples, variables, experimental methods, and conditions from the metadata of a cohort. 

Domain Experts work closely with the Data Engineer providing support on data selection and processing with their professional background. They understand the platform information and the gene measurement techniques used to determine the relevance of a cohort to the genetic question under study, help extract gene symbol information from gene annotation data, interpret or infer the patients' clinical information from the sample characteristics portion of the dataset, which is necessary for the statistical analysis.

\textbf{Statistician}
Statistician agents are assigned the task of performing regression analysis on the preprocessed datasets delivered by the Data Engineer. Its objective is to identify the genes that are predictive of the disease status
considering different conditions. To this end, two Statistician agents are designed to perform two categories of regression, i.e. single-step and two-step regression. Both types of Statistician agents will follow the instruction provided by human experts to perform the regression, then analyze the output and interpret and report the results.

\textbf{Code Reviewer}
The responsibility of the Code Reviewer agent encompasses the evaluation of code quality generated by both Statisticians and Data Engineers. At every juncture of the coding workflow, the Code Reviewer agent is tasked to review the code, as illustrated in Figure~\ref{fig:reviewer}. The process will only proceed once the code has successfully passed the review, or if the maximum number of review rounds has been reached.


\section{Criteria for manual correction of trait-condition pairs}
\label{sec:criteria}
Basically, every biomedical entity in our list can be considered a trait and paired with a condition, where the condition is either another entity in the list, or a demographic attribute ``age'' or ``gender''. However, the below rules are applied to include and exclude certain pairs to make sure that questions formed this way are scientifically valid:
\begin{itemize}
\item Entities such as language abilities, Vitamin D Levels, and bone density should only serve as the condition instead of trait;
\item Entities such as obesity and hypertension, and mental disorders like anxiety disorder and bipolar disorder should be the condition to be paired with all other traits;
\item Gender-specific entities such as prostate cancer, endometriosis, and breast cancer should not be conditioned on gender, and entities from different genders should not be paired;
\item Pairs where both the trait and condition belong to the cancer category are removed. This is because questions about genetic factors behind a cancer conditional on another type of cancer are less scientifically important.
\end{itemize}

\section{Traits and Conditions}
\label{sec:question_pairs}

\
\
\
\begin{longtable}{
  |p{\dimexpr0.25\linewidth-2\tabcolsep-\arrayrulewidth\relax}|
   m{\dimexpr0.25\linewidth-2\tabcolsep-\arrayrulewidth\relax}|
   m{\dimexpr0.5\linewidth-2\tabcolsep-\arrayrulewidth\relax}|
}
    \caption{Traits organized in 9 categories, and their corresponding conditions for the questions in our GenQEX benchmark. } \label{tab:long} \\
\hline
    \multicolumn{1}{|c|}{\centering\textbf{Type}} & \centering\textbf{Trait} & \textbf{Conditions} \\ \hline
\endfirsthead

    \multicolumn{3}{c}%
    {{\bfseries \tablename\ \thetable{} -- continued from previous page}} \\
    \hline \textbf{Type} & \textbf{Trait} & \textbf{Conditions} \\ \hline
    \endhead

    \hline \multicolumn{3}{|r|}{{Continued on next page}} \\ \hline
    \endfoot

    \hline
    \endlastfoot

    \multirow{6}{3cm}{\centering\large1. Cancer and Oncology-Related Disorders} & Liver Cancer & \small Endometriosis, Age, Hypertension, Glucocorticoid Sensitivity, Vitamin D Levels, Obesity, Susceptibility to Infections, Obstructive sleep apnea, Gender, Anxiety disorder \\ \cline{2-3}
    & Kidney Papillary Cell Carcinoma & \small Endometriosis, Age, Hypertension, Glucocorticoid Sensitivity, Vitamin D Levels, Obesity, Gender, Susceptibility to Infections, Crohn's Disease, COVID-19, Obstructive sleep apnea, Anxiety disorder \\ \cline{2-3}
    &\small Kidney Chromophobe & Gender, Age, Hypertension, Obesity, Anxiety disorder \\ \cline{2-3}
    & Stomach Cancer & \small Endometriosis, Gender, Age, Hypertension, Obesity, Anxiety disorder \\ \cline{2-3}
    & Bile Duct Cancer & \small Gender, Age, Obesity, Anxiety disorder, Hypertension \\ \cline{2-3}
    & Bladder Cancer &\small Endometriosis, Glucocorticoid Sensitivity, Gender, Age, Hypertension, Vitamin D Levels, Susceptibility to Infections, Crohn's Disease, Osteoporosis, Obstructive sleep apnea, COVID-19, Obesity, Alopecia, Anxiety disorder \\ \hline

    \multirow{1}{3cm}{\centering\large2. Cardiovascular Diseases} & Hypertension & \small Age, Breast Cancer, Lung Cancer, Gender, Prostate Cancer, Obesity, Endometriosis, Obstructive sleep apnea, Pancreatic Cancer, Vitamin D Levels, Glucocorticoid Sensitivity, COVID-19, Susceptibility to Infections, Bladder Cancer, Kidney Papillary Cell Carcinoma, Osteoporosis, Liver Cancer, Head and Neck Cancer, Esophageal Cancer, Crohn's Disease, Thyroid Cancer, Colon and Rectal Cancer, Epilepsy, Sjögren's Syndrome, Anxiety disorder \\ \hline

    \multirow{3}{3cm}{\centering\large 3. Neurological and Psychiatric Disorders} & Multiple Chemical Sensitivity &\small Anxiety disorder, Obesity, Age, Hypertension \\ \cline{2-3}
    & Anxiety disorder &\small Obstructive sleep apnea, Gender, Hypertension, Obesity, Age \\ \cline{2-3}
    & Amyotrophic Lateral Sclerosis & \small Age, Hypertension, Obesity, Gender, Anxiety disorder \\ \hline

    \multirow{3}{3cm}{\centering\large 4. Metabolic and Endocrine Disorders} & Glucocorticoid Sensitivity &\small Bladder Cancer, Pancreatic Cancer, Endometriosis, Hypertension, Lung Cancer, Breast Cancer, Prostate Cancer, Kidney Papillary Cell Carcinoma, Thyroid Cancer, Obesity, Vitamin D Levels, Crohn's Disease, Liver Cancer, Osteoporosis, Esophageal Cancer, COVID-19, Obstructive sleep apnea, Susceptibility to Infections, Colon and Rectal Cancer, Anxiety disorder \\ \cline{2-3}
    & Osteoporosis & \small Bone Density, Lung Cancer, Vitamin D Levels, Hypertension, Endometriosis, Breast Cancer, Prostate Cancer, Bladder Cancer, Age, Pancreatic Cancer, Glucocorticoid Sensitivity, Gender, Obstructive sleep apnea, Obesity, Thyroid Cancer, Psoriatic Arthritis, Head and Neck Cancer, Esophageal Cancer, Colon and Rectal Cancer, Crohn's Disease, Anxiety disorder \\ \cline{2-3}
    & Polycystic Kidney Disease & \small Gender, Hypertension, Obesity, Anxiety disorder \\ \hline

    \multirow{6}{3cm}{\large\centering 5. Genetic and Developmental Disorders} & Multiple Endocrine Neoplasia Type 2 &\small Hypertension, Anxiety disorder, Obesity \\ \cline{2-3}
    & Alopecia &\small Psoriatic Arthritis, Endometriosis, Susceptibility to Infections, Crohn's Disease, Bladder Cancer, Obesity, Hypertension, Anxiety disorder \\ \cline{2-3}
    & Intellectual Disability &\small Age, Obesity, Gender, Hypertension, Anxiety disorder \\ \cline{2-3}
    & Craniosynostosis &\small Obesity, Gender, Age \\ \cline{2-3}
    & Brugada Syndrome &\small Anxiety disorder, Hypertension, Age, Gender, Obesity \\ \cline{2-3}
    & Autoinflammatory Disorders &\small Psoriatic Arthritis, Endometriosis, Hypertension, Obesity, Anxiety disorder \\ \hline

    \multirow{2}{3cm}{\large\centering 6. Gastrointestinal and Hepatic Disorders} & Crohn's Disease &\small Susceptibility to Infections, Pancreatic Cancer, Breast Cancer, Lung Cancer, COVID-19, Bladder Cancer, Age, Glucocorticoid Sensitivity, Prostate Cancer, Endometriosis, Psoriatic Arthritis, Obesity, Celiac Disease, Sjögren's Syndrome, Vitamin D Levels, Hypertension, Alopecia, Gender, Obstructive sleep apnea, Kidney Papillary Cell Carcinoma, Thyroid Cancer, Head and Neck Cancer, Osteoporosis, Anxiety disorder \\ \cline{2-3}
    & Celiac Disease &\small Crohn's Disease, Susceptibility to Infections, Sjögren's Syndrome, Psoriatic Arthritis, COVID-19, Endometriosis, Gender, Obesity, Hypertension, Age, Anxiety disorder \\ \hline

    \multirow{2}{3cm}{\large\centering 7. Respiratory and Pulmonary Disorders} & Obstructive sleep apnea &\small Hypertension, COVID-19, Vitamin D Levels, Obesity, Endometriosis, Prostate Cancer, Bladder Cancer, Osteoporosis, Age, Breast Cancer, Lung Cancer, LDL Cholesterol Levels, Pancreatic Cancer, Susceptibility to Infections, Sjögren's Syndrome, Thyroid Cancer, Glucocorticoid Sensitivity, Crohn's Disease, Anxiety disorder, Liver Cancer, Kidney Papillary Cell Carcinoma \\ \cline{2-3}
    & COVID-19 &\small Susceptibility to Infections, Obstructive sleep apnea, Hypertension, Sjögren's Syndrome, Endometriosis, Age, Pancreatic Cancer, Crohn's Disease, Lung Cancer, Psoriatic Arthritis, Breast Cancer, Gender, Obesity, Prostate Cancer, Bladder Cancer, Vitamin D Levels, Thyroid Cancer, Head and Neck Cancer, Glucocorticoid Sensitivity, Celiac Disease, Kidney Papillary Cell Carcinoma, Anxiety disorder \\ \hline

    \multirow{2}{3cm}{\large\centering8. Rheumatological and Musculoskeletal Disorders} & Psoriatic Arthritis &\small COVID-19, Sjögren's Syndrome, Crohn's Disease, Alopecia, Autoinflammatory Disorders, Susceptibility to Infections, Celiac Disease, Osteoporosis, Obstructive sleep apnea, Gender, Age, Hypertension, Obesity, Anxiety disorder \\ \cline{2-3}
    & Sjögren's Syndrome &\small Susceptibility to Infections, COVID-19, Psoriatic Arthritis, Endometriosis, Crohn's Disease, Celiac Disease, Obstructive sleep apnea, Pancreatic Cancer, Hypertension, Thyroid Cancer, Vitamin D Levels, Breast Cancer, Age, Obesity, Anxiety disorder \\ \hline
    \multirow{1}{3cm}{\large\centering 9. Miscellaneous Traits and Conditions} & Endometriosis &\small Pancreatic Cancer, Breast Cancer, Lung Cancer, Bladder Cancer, Hypertension, Kidney Papillary Cell Carcinoma, Susceptibility to Infections, Glucocorticoid Sensitivity, Vitamin D Levels, Obesity, Head and Neck Cancer, Thyroid Cancer, Liver Cancer, COVID-19, Colon and Rectal Cancer, Esophageal Cancer, Obstructive sleep apnea, Osteoporosis, Endometrioid Cancer, Crohn's Disease, Sjögren's Syndrome, Alopecia, Stomach Cancer, Autoinflammatory Disorders, Celiac Disease, Anxiety disorder \\
    \cline{2-3}
    \multirow{3}{3cm}{} & Susceptibility to Infections &\small COVID-19, Crohn's Disease, Endometriosis, Sjögren's Syndrome, Age, Pancreatic Cancer, Lung Cancer, Hypertension, Breast Cancer, Bladder Cancer, Gender, Prostate Cancer, Obesity, Vitamin D Levels, Thyroid Cancer, Celiac Disease, Psoriatic Arthritis, Alopecia, Obstructive sleep apnea, Kidney Papillary Cell Carcinoma, Liver Cancer, Glucocorticoid Sensitivity, Esophageal Cancer, Anxiety disorder \\ \cline{2-3}
    & Kidney stones &\small Gender, Hypertension, Obesity, Anxiety disorder \\ \cline{2-3}
    & Underweight &\small Obesity, Gender, Age, Anxiety disorder, Hypertension \\
    \hline

\end{longtable}

\section{Details about the data sources}
\label{sec:datasource}
\paragraph{GEO} The Gene Expression Omnibus (GEO) \citep{clough2016gene} is a public repository that stores high-throughput gene expression data, among other types. We utilized the Entrez programming utility to systematically search the GEO database for human series data pertinent to each trait in our list, focusing on datasets with a significant sample size. Both SOFT and matrix files were downloaded for each series, with heuristic file size evaluation employed to identify datasets likely containing gene expression data. For traits yielding no results from automated searches, synonym expansion via Medical Subject Headings (MeSH) terms facilitated manual data identification.

\paragraph{TCGA-Xena} The Cancer Genome Atlas (TCGA) \citep{tomczak2015cancer}, accessible through the Xena platform \cite{goldman2020visualizing}, provides a comprehensive collection of RNAseq gene expression and clinical data across many cancer types. We extracted data for 36 traits from the TCGA cohort using the UCSC Xena platform, a repository of high-quality, cancer-related gene expression and clinical data interconnected by patient IDs.

\paragraph{NCBI Gene} The NCBI Gene database \citep{brown2015gene} serves as a vital resource for acquiring comprehensive information on gene sequences, functions, and their associations with diseases and conditions. For each trait, we queried the database to identify a set of gene symbols known to be associated with the trait, which is used for finding disease-disease associations for question generation, and selecting common regressors for two-step regression.

\end{document}